\input harvmac
\overfullrule=0pt

%
\def\sqr#1#2{{\vbox{\hrule height.#2pt\hbox{\vrule width
.#2pt height#1pt \kern#1pt\vrule width.#2pt}\hrule height.#2pt}}}
\def\Box{\mathchoice\sqr64\sqr64\sqr{4.2}3\sqr33}

\def\half{{\textstyle{1\over 2}}}
\def\hhalf{{\scriptstyle{1\over 2}}}

\Title{ \vbox{\baselineskip12pt
\hbox{hep-th/0009225}}}
{\vbox{\centerline{A Scaling Limit With} 
\bigskip
\centerline{Many Noncommutativity Parameters}}}
\smallskip
\centerline{Louise Dolan\foot{dolan@physics.unc.edu}}
\smallskip
\centerline{\it Department of Physics}
\centerline{\it
University of North Carolina, Chapel Hill, NC 27599-3255}
\bigskip
\smallskip
\centerline{Chiara R. Nappi\foot{nappi@usc.edu}} 
\smallskip
\centerline{\it Department of Physics and Astronomy}
\centerline{\it and Caltech-USC Center for Theoretical Physics}
\centerline{\it University of Southern California}
\centerline{\it Los Angeles, CA 90089-2535}

\bigskip

\noindent                 
We derive the worldsheet propagator for an open string
with different magnetic fields at the two ends, and use it to
compute two distinct noncommutativity parameters, one at each end
of the string.  The usual scaling limit that leads to noncommutative
Yang-Mills can be generalized to a scaling limit in which both
noncommutativity parameters enter.
This corresponds to expanding a theory with $U(N)$ Chan-Paton
factors around a background $U(1)^N$ gauge field with
different magnetic fields in each $U(1)$.
 
\Date{}

\lref\sw{N. Seiberg and E. Witten, ``String Theory and Noncommutative 
Geometry'', JHEP {\bf 9909} (1999) 032,  hep-th/9908142.}

\lref\acny{A. Abouelsaood, C. Callan, C. Nappi, and S. Yost,
``Open Strings in Background Gauge Fields''
Nuclear Physics {\bf B280} (1987) 599.}

\lref\an{P. Argyres and C. Nappi, 
``Spin-1 Effective Actions from Open Strings'', Nuclear Physics
{\bf B330} (1990) 151-173.}

\lref\sch{V. Schomerus, ``D-Branes and Deformation Quantization,'' JHEP
{\bf 9906} (1999) 030, hep-th/9903205.}

\lref\ft{E. Fradkin and A. Tseytlin, ``Nonlinear Electrodynamics From
Quantized Strings'', Physics Letters {\bf B163} (1985) 123.}

\lref\cp{J. Paton and H.-M. Chan, Nuclear Physics {\bf B10} (1969)
519.}   

\lref\gsw{M. Green, J. Schwarz, and  E. Witten,
{\it Superstring Theory},
vol. I and II, Cambridge University Press: Cambridge, U.K. 1987.}

\lref\polc{J. Polchinski,
{\it String Theory},
vol. I and II, Cambridge University Press: Cambridge, U.K. 1998.}

\lref\laidlaw{M. Laidlaw,
``Noncommutative Geometry from String Theory: Annulus Corrections'',
hep-th/0009068.}

\lref\ChuHo{C.-S. Chu and P.-M. Ho,
``Noncommutative Open String and D-brane'', Nucl.Phys. {\bf B550}
(1999) 151, hep-th/9812219;
``Constrained Quantization of Open String in Background B field and
Noncommutative D-brane'', Nucl.Phys. {\bf B568} 
(2000) 447, hep-th/9906192.}

\lref\Chu{C.-S. Chu, ``Noncommutative Open String: Neutral and Charged'',
hep-th/0001144.}


\newsec{Introduction and Conventions}

The propagator for the open {\it neutral} string in 
a slowly varying background $U(1)$ 
gauge field was computed in \acny\ by solving
the equations of motion
\eqn\emo{\Box \,< x^i(z,\bar z) \, x^j(\zeta,\bar \zeta) > \,
= - 2\pi\alpha' \delta^2(z-\zeta)\, G^{ij}}
with the boundary conditions
\eqn\bcnp{(\partial_z -\partial_{\bar z})
< x^i(z,\bar z) \, x^j(\zeta,\bar \zeta) > \, 
+  2\pi\alpha' \,g^{ik} B_{k\ell} (\partial_z +\partial_{\bar z})\,
< x^\ell (z,\bar z) \, x^j(\zeta,\bar \zeta) >\,|_{z=\bar z} \, = 0}
where  $\Box \equiv 4\partial_z\partial_{\bar z}$.
Here the open string worldsheet $\Sigma$ will denote the disk
with Euclidean metric $\gamma^{\alpha\beta} = \delta^{\alpha\beta}$.
The complex coordinates $z, \bar z$ are related to the 
original strip coordinates $\sigma,\tau$ with $\tau$ rotated to 
$t\equiv i\tau$ and 
$-\infty\le t\le\infty \,, 0\le \sigma \le \pi\,,\,$  
by $z\equiv e^{t + i\sigma}\,, \bar z\equiv e^{t - i\sigma}$ 
with ${\rm Im} z \ge 0$.

The propagator was found to be\foot{We have adopted here 
the rewriting of \sw, which corresponds to choosing  a particular definition 
for the normal ordering of 
the  position zero mode $x_0^i$ which appears in the
normal mode expansion of $x^i(z,\bar z)$, namely  $:x_0^i x_0^j:
\equiv \half (x_0^i x_0^j + x_0^j x_0^i).$}  

\eqn\anprop{\eqalign{< x^i(z,\bar z) \, x^j(\zeta,\bar \zeta) >
&= - \alpha' \,[\half g^{ij}  \ln (z - \zeta)
+ \half g^{ij}  \ln (\bar z - \bar\zeta)\cr
&\hskip30pt + ( -\half g^{ij} + G^{ij} + {\theta^{ij}\over
{2\pi\alpha'}} ) \ln (z - \bar\zeta)\cr
&\hskip30pt + ( -\half g^{ij} + G^{ij} - {\theta^{ij}\over
{2\pi\alpha'}} ) \ln (\bar z - \zeta)\,\,
-{i\over{2\alpha'}} \theta^{ij}\,]\cr}} 
where
\eqn\geeig{\eqalign{G^{ij} & = [ (g + 2\pi\alpha' B)^{-1}\, g \, 
(g -  2\pi\alpha' B)^{-1}]^{ij} \cr
G_{ij}  &\equiv  g_{ij} - (2\pi\alpha')^2 ( B g^{-1} B)_{ij} 
=  [ (g - 2\pi\alpha' B)\, g^{-1} \,
(g + 2\pi\alpha' B)]_{ij}\cr
\theta^{ij} &\equiv - (2\pi\alpha')^2 [ (g + 2\pi\alpha' B)^{-1}\, B \,
(g -  2\pi\alpha' B)^{-1}]^{ij}\,. \cr}}  

In \sw\ the above propagator was used to compute the ``equal time'' commutator
of the string operators via a short distance expansion procedure \sch\ ,
and to define a noncommutativity parameter. This parameter entered in 
the definition
of the star product of the noncommutative  gauge theory  derived in the 
scaling limit. The limiting gauge theory could be abelian or non-abelian 
(the latter case could be achieved by introducing Chan-Paton matrices
\refs{\cp,\gsw,\polc} in 
string amplitudes), but it only involved a single noncommutative parameter.
This scaling limit described
{\it massless} charged and neutral non-abelian gluons 
living on a
noncommutative geometry with one noncommutative parameter $\theta^{ij}$. 

In this paper we  start with a string 
theory in a non-abelian $U(N)$ background and study its scaling limit.
Actually, we will restrict ourselves to backgrounds that reside in the 
$U(1)^N$ Cartan subgroup, but with different constant background $U(1)$ fields
on each brane. The novelty compared to the $U(1)$ case is the presence of
 charged strings.  

The first step will be to
compute the propagator on the disk, which we do  
by starting with the mode expansion
for the charged string originally derived in \acny\ .
Hence, in sect. 2 we will rederive it in a more convenient notation, 
and pay 
special attention to the zero modes 
so to ensure that the the charged string mode expansion converges to the 
neutral one when the background fields at the two ends become the same. 
In sect. 4, we
will then use this mode expansion  
together with the commutation relations to 
compute the charged string propagator. 
We will have to pay attention to which  
string states we use to define the propagator, 
as we want to recover the usual neutral string propagator \anprop\ ,
when the backgrounds are the same at the two ends.  
This will require evaluating the propagator
between coherent states of the Landau levels.
In sect. 3, we therefore discuss the spectrum of the charged string.

In sect. 5, we repeat the usual argument \sch\ , but now compute two 
noncommutativity parameters, one at each end of the string. 
This emphasizes the interpretation that the noncommutativity of the
$D$-brane worldvolume in the presence of a background $B$-field along the
brane is really
a property of the endpoint of the string, rather than a feature of the
worldvolume itself. For a $U(1)^N$ background, there are $N$ different
noncommutativity parameters.
In sect.6, we compute the short distance behavior of the operator products
of tachyon vertex operators which are inserted on the boundaries, 
and show they reduce to star products, 
with two different noncommutativity parameters.
These enter into the computation of the scattering amplitudes of the
scaling limit theory.

In sect.7, we discuss the spectrum of the theory and show that there is a scaling limit in this more general background.
In addition to the $U(1)^N$ massless gauge bosons, 
charged vector states survive for each Landau level.
As in the string theory we start from,
the states of the limiting non-abelian
noncommutative gauge theory are no longer massless, but rather
tachyonic or massive.  An obvious question is what is the generalization
 of the $\hat {F^2}$ noncommutative Lagrangian. 
Our construction 
provides an example of the bimodules discussed in \sw\ . 

\newsec{Charged Open String Normal Mode Expansion}

The complete worldsheet action for the ``charged'' string
with different magnetic fields at each end is 
\eqn\wsacttwo{\eqalign{S &= {1\over{4\pi\alpha'}}\int_{\Sigma}
g_{\mu\nu}\partial_\alpha X^\mu \,\partial^\alpha X^\nu   
-{i\over 2} \int_{-\infty}^{\infty} dt \, 
(B^{(1)}_{ij} X^i \partial X^j\,|_{\sigma= 0}
+ \, B^{(2)}_{ij} X^i \partial X^j\,|_{\sigma= \pi} \,)\,.\cr}}  
Here $0\le\mu,\nu\le 25\,$
and the open strings end on $D p$-branes in the $(0,i)$ 
directions for $1\le i\le p$.
Variation of \wsacttwo\ gives the equations of motion 
for the worldsheet field
\eqn\eomx{(\partial_\sigma^2 + \partial_t^2) \,X^\mu(\sigma, t) = 0}
and the boundary conditions at each end of the string are
\eqn\bcx{\eqalign{&\hskip-55pt g_{ij} \partial_\sigma X^j +
2\pi i\alpha' B^{(1)}_{ij} \partial_t X^j \,|_{\sigma = 0} = 0\cr
&\hskip-55pt g_{ij} \partial_\sigma X^j -
2\pi i\alpha' B^{(2)}_{ij} \partial_t X^j \,|_{\sigma = \pi} = 0\cr}}
\eqn\bco{\eqalign{&\hskip20pt\partial_\sigma X^0 \,|_{\sigma =0, 
\sigma = \pi} = 0\,\cr 
&\hskip20pt X^I \,|_{\sigma =0, \sigma = \pi} = 0\,,\quad p+1\le I\le 25\,.}}
(Note that the worldsheet action for the neutral string, 
whose propagator in the directions along the $D p$-brane
is given in \anprop\ ,
is a special case of \wsacttwo\ with $B^{(1)}_{ij}=-B^{(2)}_{ij}$).

For simplicity, we 
now specialize to the case where the ends of the string
live on  $D2$-branes, 
{\it i.e.} $i=1,2$. In this case the magnetic fields have only one
component and we relabel them as
$B^{(1)}_{12} = q_1 B_{12}$ and $B^{(2)}_{12} = q_2 B_{12}$,
where $q_1 + q_2 \ne 0$.
We choose a diagonal metric in the $1,2$ directions
with equal components $g_{ij}\equiv g^{-1} \delta_{ij}$, and  
retain an overall factor since eventually we will scale the metric.
The open string metric defined in (2.8) is thus also diagonal
$G_{ij} = \delta_{ij} G^{-1}$.

In a basis given by
$X^\pm(\sigma, t) \equiv  X^1 (\sigma, t) \pm i  X^2 (\sigma, t)$,
the charged string normal mode expansion \acny,\an\ can be written as
\eqn\nmp{ X^+ (z,\bar z) = x^+ 
+ {\textstyle {i\over 2}}\sqrt{2\alpha'}\sum_{r\in {\cal Z} + A}
{a_r\over r} ( z^{-r} + \bar z^{-r}) 
-{\textstyle {1\over 2}}\sqrt{2\alpha'} B \sum_{r\in {\cal Z} + A}
{a_r\over r} ( z^{-r} - \bar z^{-r} )} 
\eqn\nmm{ X^- (z,\bar z) = x^-
+ {\textstyle {i\over 2}}\sqrt{2\alpha'}\sum_{s\in {\cal Z} - A}
{\tilde a_s\over s} ( z^{-s} + \bar z^{-s})
+{\textstyle {1\over 2}}\sqrt{2\alpha'} B \sum_{s\in {\cal Z} - A}
{\tilde a_s\over s} ( z^{-s} - \bar z^{-s} )\,.}  
with  commutation relations\foot{The $[x^+, x^-]$ commutation relation, which
provided the first indication of noncommutativity
of spacetime in the context of strings in background gauge fields, was
derived in \acny\ ,  and re-examined in \ChuHo\ and \Chu\ .}  
\eqn\cscr{\eqalign{[ a_r , \tilde a_s ] &= 
2 G \,r \,\delta_{r,-s}\,;\qquad
[ a_r , a_{r'} ] = 0 = [ \tilde a_s , \tilde a_{s'} ]\,;\cr
[ x^+, x^- ] &=  - {2 \over {(q_1 + q_2) B_{12}}}\,;\qquad
[ a_r , x^{\pm} ] = 0 = [ \tilde  a_s , x^{\pm} ]\,;\cr}} 
where
\eqn\somedef{\eqalign{&G = {g\over{1+B^2}}\cr
&B \equiv g q_1 2\pi\alpha' B_{12}\,,\cr
&A ={1\over \pi} ( \arctan B + \arctan {q_2\over q_1} B )\,.\cr}}
The oscillators $a_r,\tilde a_s$ are non-integrally moded 
with $r = n+A$,  $s = n-A$, for $n\in {\cal Z}$. 
The operators have hermiticity 
$a_r^\dagger = \tilde a_{-r}$ and $(x^+)^\dagger = x^-$. 
The comparison with the neutral string case is given by the limit 
as $A\rightarrow 0$, and it 
requires care with the zero modes:
\eqn\nmpn{\eqalign{\lim_{A\rightarrow 0}  X^+ (z,\bar z) &= 
\lim_{A\rightarrow 0}\, ( x^+ + i \sqrt{2\alpha'} {a_A\over A} ) 
\,- {\textstyle {i\over 2}}\sqrt{2\alpha'} a_0 \,\ln z\bar z    
\,+ {\textstyle {1\over 2}}\sqrt{2\alpha'} 
B \, a_0 \,\,\ln {z\over {\bar z}}\cr
&\hskip10pt +  {\textstyle {i\over 2}}\sqrt{2\alpha'} \sum_{n\ne 0}
{a_n\over n} ( z^{-n} + \bar z^{-n}) 
-{\textstyle {1\over 2}}\sqrt{2\alpha'} B \sum_{n\ne 0}
{a_n\over n} ( z^{-n} - \bar z^{-n} )}}                            
\eqn\nmmn{\eqalign{\lim_{A\rightarrow 0}  X^- (z,\bar z) &=
\lim_{A\rightarrow 0}\, ( x^- - i \sqrt{2\alpha'} {{\tilde a_{-A}}\over A} )
\,- {\textstyle {i\over 2}}\sqrt{2\alpha'} {\tilde a_0} \,\ln z\bar z
\,- {\textstyle {1\over 2}}\sqrt{2\alpha'} B \,  {\tilde a_0} 
\,\,\ln {z\over {\bar z}}\cr
&\hskip10pt +  {\textstyle {i\over 2}}\sqrt{2\alpha'} \sum_{n\ne 0 }
{\tilde a_n\over n} ( z^{-n} + \bar z^{-n})
+{\textstyle {1\over 2}}\sqrt{2\alpha'} B \sum_{n\ne 0}
{\tilde a_n\over n} ( z^{-n} - \bar z^{-n} )\,.}}

To make contact with the neutral string zero mode operators, we need
therefore to identify the limits
\eqn\zeromdef{\eqalign{\lim_{A\rightarrow 0} \,\,
( x^+ + i\sqrt{2\alpha'} {a_A\over A}) &= x^+_0\cr 
\lim_{A\rightarrow 0} \,\,(x^- - i\sqrt{2\alpha'} {a_{-A}\over A})
&= x^-_0\,.\cr}}
The neutral string commutation relations in the $\pm$ basis are 
\eqn\nscr{\eqalign{[ a_n , 
\tilde a_m ] &= 2 G \,n \,\delta_{n,-m}\,;\qquad
[ a_n , a_{n'} ] = 0 = [ \tilde a_n , \tilde a_{n'} ]\,;\cr
[ x^+_0, x^-_0 ] &= 2 \,\Theta^{12}\,; \qquad
[ a_n , x^{\pm}_0 ] = 0 = [ \tilde  a_n , x^{\pm}_0 ]\,\,{\rm for}\,\, n\ne 0\,
;\cr
[ x^+_0, \tilde a_0 ] &= i \sqrt{2\alpha'} \,2 G =
[ x^-_0, a_0 ]\,; \qquad [ x^+_0, a_0 ] = 0 =  [ x^-_0, \tilde a_0 ]\,\cr}}
with $\Theta^{12} = -2\pi\alpha' B G$.
Indeed using
\eqn\limit{\eqalign{\lim_{A\rightarrow 0} (\pi A)^{-1}
&=\lim_{(Q=q_1+q_2)\rightarrow 0} [\arctan ({q_1^{-1} B Q\over 
{1 +B^2 - q_1^{-1} Q B^2}}) \,]^{-1}\cr
&\sim 
({{1+B^2}\over {q_1^{-1} B Q}}) ( 1 - {{q_1^{-1} Q B^2}\over {1+B^2}})\,.}}

one can show that the identification in \zeromdef\ is consistent since
\eqn\zeromcr{\eqalign{ &  \lim_{A\rightarrow 0} \,\, 
[  x^+ + i\sqrt{2\alpha'} {a_A\over A} , 
x^- - i\sqrt{2\alpha'} {\tilde a_{-A}\over A}\,]\cr
&=  \lim_{A\rightarrow 0} \,\,
( [  x^+, x^- ] + 2\alpha' {1\over {A^2}} [ a_A, \tilde a_{-A} ] )\cr
&=  \lim_{A\rightarrow 0} \,\,
({ - 4\alpha' g \pi\over {B + {q_2\over q_1}B}}\, 
+ 4 \alpha' \,{1\over A} \,G )\cr
&= { - 4\alpha' g B \pi\over {1 + B^2}} = 2 \, \Theta^{12}\,,\cr}} 
and $[ x^+_0, x^-_0 ] = 2 \,\Theta^{12}$ from \nscr\ .

\newsec{Charged String Spectrum and Coherent States}

We briefly review the charged string spectrum\acny\ in our notation, 
so that we can study its scaling limit in sect.7.
We will also consider coherent states in order to  
introduce the states we use to define the charged string propagator.  
The commutation relations \cscr\ correspond to the operator product
expansion
\eqn\opevvo{\eqalign{a(z)\,\tilde a(\zeta) &=
(z-\zeta)^{-2} \,2 G \,[ A \,({\textstyle{\zeta\over z}})^{A-1}
+ (1-A) \,({\textstyle{\zeta\over z}})^A ]\cr
&\hskip10pt + \,{\rm N} \,a(z) \,\tilde a(\zeta) \cr}}
where we define complex worldsheet bosons 
$a(z) = \sum_{r\in {\cal Z} + A} a_r z^{-r-1}$,
$\tilde a(z) = \sum_{s\in {\cal Z} - A} \tilde a_s z^{-s-1}$, 
and the normal ordering as 
\eqn\no{\eqalign{{\rm N}  \,a_r \,\tilde a_s \equiv \,: a_r \,\tilde a_s :\,
\equiv \,&\tilde a_s\,a_r\quad
{\rm for}\, r\ge A\,,\, s\le -A\,,\cr
&a_r \,\tilde a_s\quad  {\rm otherwise}\,.\cr}}
The Virasoro current is 
$L(z) = \hhalf G^{-1} : a(z)\,\tilde a(z) : 
+ z^{-2} \hhalf A (1-A)$
with central charge equal to 2 corresponding to the two bosonic degrees of 
freedom. The commutation relations with the oscillators are
\eqn\comrel{[L_n, a_r ] = (n-r) \,a_{r+n}\,,\qquad
[L_n, \tilde a_s ] = (n-s) \,\tilde a_{s+n}\,.}
As in \acny\ we take the states to be eigenstates of 
the position operator $x^+$ and
label them with the continuous eigenvalue $x_+$ as
$|x_+\rangle\equiv e^{ (-\half (q_1 + q_2) B_{12} x_+) x^- }
|0\rangle $, since from \cscr\ we have
$ x^+ |x_+\rangle = x_+ |x_+\rangle$. Notice that we have  
$a_r | x_+\rangle = 0$ for $r\ge A$, and
$\tilde a_s | x_+\rangle = 0$ for $s\ge 1-A$. In \somedef\ 
it is sufficient to consider $0\le A\le\half $. 
Then \eqn\loonx{L_0 |x_+\rangle = \hhalf A (1-A) |x_+\rangle}
for each of the infinite number of states $|x_+\rangle$. So they 
are degenerate all with the same tachyonic mass 
\eqn\masscon{\alpha' m^2 = -1 +  \hhalf A (1-A) < 0}
since the contribution from these degrees of freedom
to the mass operator is $\alpha' m^2 = L_0 -1$.
There is a tower of oscillator states built from $| x_+\rangle$
starting with 
\eqn\tower{a_{-1+A} | x_+\rangle\,,\qquad 
\tilde a_{-1-A} | x_+\rangle} which have masses
$\alpha' m^2 = -\hhalf A (1+A) < 0$ and 
$\alpha' m^2 = \hhalf A (3-A)>0 $ respectively.
Although $a_A | x_+\rangle = 0$,   
we can also consider the set of states
\eqn\landau{ \tilde a_{-A} | x_+\rangle \,,\quad
\tilde a_{-A} \tilde a_{-A} | x_+\rangle \,,\qquad {\rm etc.}}
with masses $ -1 +  \hhalf A (1-A) + A$;
$ -1 +  \hhalf A (1-A) + 2A$, etc. 
The states in \landau\ together with $ | x_+\rangle $
are separated from each other in $\alpha' m^2$ values by $A$.
This is reminiscent of the equally spaced Landau levels with 
frequency separation $q B$ of a charged
particle in a constant magnetic field where here $A$ plays the
role of $q B$;
so $| x_+\rangle, \tilde a_{-A} | x_+\rangle, 
\tilde a_{-A} \tilde a_{-A} | x_+\rangle,$
etc. form equally spaced Landau levels each of infinite degeneracy
and each is the lowest state of an oscillator tower such as that described 
in \tower\ . 

To compute the charged string propagator we introduce the states
\eqn\alp{ |\,\alpha \,\rangle \equiv
e^{- (\half (q_1 + q_2) B_{12})\, x^-}
\, | 0\rangle}
\eqn\bet{ \langle \,\beta\,| \equiv \langle  0 | \,
e^{-\half ( x^+ + {i\sqrt{2\alpha'}\over A} a_A )}}
since they correspond in the $A\rightarrow 0$ limit to the vacuum states
defined by the normal ordering $N x^+_0 x^-_0 \equiv
\half (  x^+_0 x^-_0 +  x^-_0 x^+_0 )$ in the neutral string case,
 {\it i.e.}  these states have the property
\eqn\states{\eqalign{
& \lim_{A\rightarrow 0} \langle \,\beta\,|
( x^+ + {i\sqrt{2\alpha'}\over A} a_A )
( x^- - {i\sqrt{2\alpha'}\over A} \tilde a_{-A} )
|\,\alpha\,\rangle \cr
&= \lim_{A\rightarrow 0} \,(\,[ x^+, x^- ] + {2\alpha'\over A^2}
[ a_A , \tilde a_{-A} ] +
\langle \,\beta\,| ( x^- - {i\sqrt{2\alpha'}\over A} \tilde a_{-A} )
\, x^+ |\,\alpha\,\rangle \,)\cr
&=  2 \Theta^{12} -\half
\lim_{A\rightarrow 0} ({-4\pi\alpha' g\over { B + {q_2\over q_1}B}}
+ {4 \alpha' G\over A})\cr
&= \Theta^{12}
=\lim_{A\rightarrow 0} \langle \,\beta\,| x^+_0 \, x^-_0
|\,\alpha\,\rangle\,,\cr}}
and
\eqn\statestwo{\eqalign{
& \lim_{A\rightarrow 0} \langle \,\beta\,|
( x^- - {i\sqrt{2\alpha'}\over A} \tilde a_{-A} )
( x^+ +  {i\sqrt{2\alpha'}\over A} a_{A} )
|\,\alpha\,\rangle \cr      
&= - \Theta^{12}  
=\lim_{A\rightarrow 0} \langle \,\beta\,| x^-_0 \, x^+_0
|\,\alpha\,\rangle\,.\cr}}
The states have been normalized
$ \langle \,\beta\,|\,\alpha\,\rangle =
1 \,.$
Note that the state 
$\langle \beta \,|$ is a coherent state made up from the 
Landau level $\tilde a_{-A} \,| \,  0\rangle $.        

\newsec{Charged String Propagator}

In order to compute a charged propagator that reduces to the
neutral expression \anprop\ as $A$ goes to zero, we consider 
the charged string propagator on the disk evaluated between the states
$|\alpha\rangle$ and $|\beta\rangle$.
For $|z|> |\zeta|$, we find it is given by
\eqn\pmprop{\eqalign{< X^+ (z,\bar z) X^- (\zeta,\bar\zeta) > &\equiv 
\langle \beta\, | X^+ (z,\bar z) X^- (\zeta,\bar\zeta) |\,\alpha\, \rangle\cr
&= {-2\alpha'\pi g\over {B + {q_2\over q_1} B}}
- 2\alpha' G {\textstyle{1\over A}} (\zeta^A + \bar \zeta^A - 1)
+ 2 i \alpha' G B {\textstyle{1\over A}} (\zeta^A - \bar \zeta^A)\cr
&\hskip10pt +\alpha' G \quad[ f({\textstyle{{\zeta\over z}}}) +  
f({\textstyle{\bar\zeta\over {\bar z}}})
+ f({\textstyle{\zeta\over {\bar z}}}) 
+ f({\textstyle{\bar\zeta\over z}}) ]\cr
&\hskip10pt +\alpha' G \, B^2 \quad[ f({\textstyle{{\zeta\over z}}}) +
f({\textstyle{\bar\zeta\over {\bar z}}})
- f({\textstyle{\zeta\over {\bar z}}})
- f({\textstyle{\bar\zeta\over z}}) ]\cr 
&\hskip10pt + 2 i\alpha' G \, B \quad
[ - f({\textstyle{\zeta\over {\bar z}}})
+ f({\textstyle{\bar\zeta\over z}}) ]\cr}} 
where 
\eqn\defin{\eqalign{f(\rho) \equiv &\sum_{r=n+A;\, n\ge 0} {\rho^r\over r}\,;
\qquad\quad \lim_{A\rightarrow 0} f(\rho) = - \ln (1 - \rho)\ +  
\lim_{A\rightarrow 0} {\rho^A\over A}\,.\cr}}

The other non-zero component of the 
charged string propagator, for $|z|> |\zeta|$, is
\eqn\mpprop{\eqalign{ < X^- (z,\bar z) X^+ (\zeta,\bar\zeta) > &\equiv
\langle \beta \,|  X^- (z,\bar z) X^+ (\zeta,\bar\zeta) |\,\alpha\,\rangle\,\cr
&=  {2\alpha'\pi g\over {B + {q_2\over q_1} B}}
- 2\alpha' G {\textstyle{1\over A}} (z^A + \bar z^A - 1)
+ 2 i \alpha' G B {\textstyle{1\over A}} (z^A - \bar z^A)\cr
&\hskip10pt +\alpha' G \quad[ g({\textstyle{{\zeta\over z}}}) +
g({\textstyle{\bar\zeta\over {\bar z}}})
+ g({\textstyle{\zeta\over {\bar z}}})
+ g({\textstyle{\bar\zeta\over z}}) ]\cr
&\hskip10pt +\alpha' G \, B^2 \quad[ g({\textstyle{{\zeta\over z}}}) +
g({\textstyle{\bar\zeta\over {\bar z}}})          
- g({\textstyle{\zeta\over {\bar z}}})
- g({\textstyle{\bar\zeta\over z}}) ]\cr
&\hskip10pt - 2 i\alpha' G \, B \quad
[ - g({\textstyle{\zeta\over {\bar z}}})
+ g({\textstyle{\bar\zeta\over z}}) ]\cr}}
and
\eqn\defing
{\eqalign{ g(\rho) \equiv &\sum_{s=n-A;\, n\ge 1} {\rho^s\over s}\,;
\qquad\quad \lim_{A\rightarrow 0} g(\rho) = - \ln (1 - \rho)\,.\cr}}  
To compute \pmprop,\mpprop\ we used the normal mode expansion
\nmp,\nmm, the commutation relations \cscr, the
normal ordering defined in \no, and the states $|\alpha\rangle,
|\beta\rangle$ described in \alp, \bet .
For all $z,\zeta$, the charged string propagator is
\eqn\pmpropall{\eqalign{&G^{+-}(z,\bar z; \zeta, \bar\zeta)\equiv
\langle \beta \,| T X^+ (z,\bar z) X^- (\zeta,\bar\zeta) 
|\,\alpha\, \rangle\cr
&= \theta ( |z| - |\zeta | ) \,\cdot \, <X^+ (z,\bar z) 
X^- (\zeta,\bar\zeta) > 
+ \theta ( |\zeta | - |z| )
\,\cdot \, < X^- (\zeta,\bar\zeta)  X^+ (z,\bar z) >\,. \cr}}
The other component is given from the symmetry property
\eqn\symcs{ G^{-+}(z,\bar z; \zeta, \bar\zeta )
= G^{+-}(\zeta, \bar\zeta ; z,\bar z)\,.}
These are the propagators that reduce to the neutral string expressions 
since we can show
\eqn\limcsproppm{\eqalign{\lim_{A\rightarrow 0}
G^{+-}(z,\bar z; \zeta, \bar\zeta)
&=  < x^+ (z,\bar z) x^- (\zeta,\bar\zeta) >\cr}}
\eqn\limcspropmp{\eqalign{\lim_{A\rightarrow 0}
G^{-+}(z,\bar z; \zeta, \bar\zeta)
&=  < x^- (z,\bar z) x^+ (\zeta,\bar\zeta) >\,.\cr}}
The right hand sides in 
\limcsproppm, \limcspropmp\ are equivalent to the neutral propagators 
\anprop\ in the $\pm$ basis.
For completeness, we mention that other limits are also consistent.
For example, as $\zeta\rightarrow 0$, we have  
\eqn\cszero{\eqalign{ G^{+-}(z,\bar z; 0) &=
{-2\alpha'\pi g\over {B + {q_2\over a_1} B}} + {2\alpha' G\over A}
\,,{\rm for} A > 0\,;\cr
G^{-+}(z,\bar z; 0) &= {2\alpha'\pi g\over {B + {q_2\over a_1} B}}
- 2\alpha' G  {\textstyle{1\over A}} (z^A + \bar z^A - 1)
+ 2 i \alpha' G B {\textstyle{1\over A}} (z^A - \bar z^A)\,.\cr}}
If we consider the simultaneous limit $\zeta\rightarrow 0$, $A\rightarrow 0$,
in the expression for $ G^{+-}(z,\bar z;\zeta,\bar\zeta)$ we must
be careful to use
${\lim_{\zeta\rightarrow 0\atop {A\rightarrow 0}}
\zeta^A = \lim_{\zeta\rightarrow 0\atop {A\rightarrow 0}} e^{A\ln\zeta}
= \lim_{x\rightarrow 0} e^{x\ln x} = 1\,. }$ Then
\eqn\limnsz{\eqalign{\lim_{\zeta\rightarrow 0\atop{A\rightarrow 0}}
G^{+-}(z,\bar z; \zeta,\bar\zeta )
&= < x^+ (z,\bar z) x^- (0) >\cr
\lim_{A\rightarrow 0} G^{-+}(z,\bar z; 0)
&=  < x^- (z,\bar z) x^+ (0) >\,.\cr}}

\newsec{The Set of Noncommutativity Parameters}

We are going to compute the ``equal time'' commutator of the 
string operators at coincident points on the boundary \sch,\sw\ . 
We distinguish the two boundary regions of the open string disk as follows.
On the boundary $\sigma = 0$, we have $z = |z| = \tau$ and
$\zeta = |\zeta | = \tau'$ so $\tau,\tau'>0$; while
on $\sigma = \pi$, then $z = |z| e^{i\pi}= \tau$ and
$\zeta = |\zeta |  e^{i\pi} = \tau'$ so here $\tau,\tau'<0$.
We will evaluate both of the propagators on the boundaries $\sigma = 0$ and
$\sigma = \pi$, 
and find a different noncommutativity 
parameter\foot{After completion of our paper a preprint \laidlaw\
appeared which derives these parameters from a charged string annulus
propagator. Since the result is a short distance effect, it is
independent of the topology of the worldsheet.}
at each end of the string.
For $|z|> |\zeta|$, and on the boundary $\sigma = 0$, we have
\eqn\bzero{\eqalign{< X^+ (z,\bar z) X^- (\zeta,\bar\zeta) > |_{\sigma = 0}
&= {-2\alpha'\pi g\over {B + {q_2\over q_1} B}}\quad
+ 4\alpha' G \sum_{n=0}^\infty {1\over {n+A}} 
({\textstyle{{\zeta\over z}}})^{n+A} \cr
&\hskip15pt + 
{2\alpha' G \over A} - {4\alpha' G \over A}\,\zeta^A\,,\cr
< X^- (z,\bar z) X^+ (\zeta,\bar\zeta) > |_{\sigma = 0}
&=  {2\alpha'\pi g\over {B + {q_2\over q_1} B}}\quad 
+ 4\alpha' G \sum_{n=1}^\infty {1\over {n-A}}
({\textstyle{{\zeta\over z}}})^{n-A}\cr
&\hskip15pt + {2\alpha' G \over A} - {4\alpha' G \over A}
\, z^A\,.\cr}}
We now compute the commutator that the defines the 
noncommutativity parameter at $\sigma = 0$.

\eqn\scho{\eqalign{[ X^+ (\tau) ,  X^- (\tau)] 
&= T (  X^+ (\tau) \, X^- (\tau^-) -  X^+ (\tau)  X^- (\tau^+) )\cr
&\equiv \lim_{\epsilon\rightarrow 0}
(  < X^+ (\tau) \, X^- (\tau -\epsilon ) > - 
< X^- (\tau + \epsilon )\, X^+ (\tau) > ) 
\, , \quad ({\rm for}\, \epsilon > 0)\cr
&=  \lim_{\epsilon\rightarrow 0} \,
( \quad {-4\alpha'\pi g\over {B + {q_2\over q_1} B}}\quad 
+  4\alpha' G [ \sum_{n=0}^\infty {1\over {n+A}}
({\textstyle{{\tau - \epsilon \over \tau }}})^{n+A}
\, -  \sum_{n=1}^\infty {1\over {n-A}}
({\textstyle{{\tau \over \tau + \epsilon }}})^{n-A} \,] \cr
&\hskip35pt - {4\alpha' G\over A} (\tau -\epsilon )^A
+ {4\alpha' G\over A} (\tau +\epsilon )^A \quad )\cr 
&= {-4\alpha'\pi g\over {B + {q_2\over q_1} B}}\quad
+  4\alpha' G \,\pi \cot {\pi A} \cr  
&=  -4\alpha'\pi (g)^2 {q_1 2\pi\alpha' B_{12}\over 
{1 +  (g)^2 (q_1 2\pi\alpha' B_{12})^2}}\cr
&= 2 \Theta ^{12}\,.\cr}}
Notice that $\Theta^{12}$ is the same expression that appears
in the neutral string commutation relations \nscr .
For $|z|> |\zeta|$, and on the boundary $\sigma = \pi $, we have
\eqn\bpi{\eqalign{< X^+ (z,\bar z) X^- (\zeta,\bar\zeta) > |_{\sigma = \pi}
&= {-2\alpha'\pi g\over {B + {q_2\over q_1} B}}
+ {2\alpha' G \over A}\cr
&\hskip10pt + 2\alpha' G ( 1 + B^2 ) \sum_{n=0}^\infty {1\over {n+A}}
({\textstyle{{|\zeta|\over |z| }}})^{n+A} \cr
&\hskip10pt + 2\alpha' G ( 1 - B^2 ) \cos {2\pi A} \,
\sum_{n=0}^\infty {1\over {n+A}}
({\textstyle{{|\zeta|\over |z| }}})^{n+A} \cr
&\hskip10pt + i 2\alpha' G B (- 2i \sin {2\pi A} ) \,
\sum_{n=0}^\infty {1\over {n+A}}
({\textstyle{{|\zeta|\over |z| }}})^{n+A} \cr
&\hskip10pt - {4\alpha' G \over A} |\zeta|^A \cos{\pi A} 
-  {2\alpha' G B \over A}  |\zeta|^A \sin{\pi A}\,.\cr }}
\eqn\bpim{\eqalign{< X^- (z,\bar z) X^+ (\zeta,\bar\zeta) > |_{\sigma = \pi}
&=  {2\alpha'\pi g^{11}\over {B + {q_2\over q_1} B}}
+ {2\alpha' G \over A}\cr
&\hskip10pt + 2\alpha' G ( 1 + B^2 ) \sum_{n=1}^\infty {1\over {n-A}}
({\textstyle{{|\zeta|\over |z| }}})^{n-A} \cr
&\hskip10pt + 2\alpha' G ( 1 - B^2 ) \cos {2\pi A} \,
\sum_{n=0}^\infty {1\over {n-A}}
({\textstyle{{|\zeta|\over |z| }}})^{n-A} \cr
&\hskip10pt + i 2\alpha' G B ( 2i \sin {2\pi A} ) \,
\sum_{n=0}^\infty {1\over {n-A}}
({\textstyle{{|\zeta|\over |z| }}})^{n-A} \cr
&\hskip10pt  - {4\alpha' G \over A} |z|^A \cos{\pi A}
-  {2\alpha' G B \over A} |z|^A \sin{\pi A}\,.\cr }} 
\vfill\eject

The noncommutativity parameter at the $\sigma = \pi$ end of the string
is defined from the propagators on the boundary $\tau = {\rm Re} z <0,
\tau' = {\rm Re} \zeta <0$,to be
\eqn\schpi{\eqalign{[ X^+ (\tau) ,  X^- (\tau)]
&= T (  X^+ (\tau) \, X^- (\tau^-) -  X^+ (\tau)  X^- (\tau^+) )\cr
&=\equiv  \lim_{\epsilon\rightarrow 0}
(  < X^+ (\tau) \, X^- (\tau +\epsilon ) > -
< X^- (\tau - \epsilon )\, X^+ (\tau) > )
\, , \quad ({\rm for}\, \epsilon > 0)\cr
&=  \lim_{\epsilon\rightarrow 0} \,
( \quad {-4\alpha'\pi g\over {B + {q_2\over q_1} B}}\cr
&\hskip10pt +  2\alpha' G ( 1 + B^2 ) [ \sum_{n=0}^\infty {1\over {n+A}}
({\textstyle{{|\tau + \epsilon| \over |\tau | }}})^{n+A}   
\, -  \sum_{n=1}^\infty {1\over {n-A}}
({\textstyle{{|\tau | \over |\tau -\epsilon |}}})^{n-A} \,]\cr
&\hskip10pt + 2\alpha' G ( 1 - B^2 ) ( 1 - 2 \sin^2 \pi A )\,
[ \sum_{n=0}^\infty {1\over {n+A}}
({\textstyle{{|\tau + \epsilon| \over |\tau | }}})^{n+A}
\, -  \sum_{n=1}^\infty {1\over {n-A}}
({\textstyle{{|\tau | \over |\tau -\epsilon |}}})^{n-A} \,]\cr
&\hskip10pt  + i 2\alpha' G  B ( - 2i\sin 2\pi A )\,
[ \sum_{n=0}^\infty {1\over {n+A}}
({\textstyle{{|\tau + \epsilon| \over |\tau | }}})^{n+A}
\, -  \sum_{n=1}^\infty {1\over {n-A}}
({\textstyle{{|\tau | \over |\tau -\epsilon |}}})^{n-A} \,]\cr 
&\hskip10pt  - {4\alpha' G \over A} |\tau + \epsilon |^A \cos{\pi A}
-  {2\alpha' G B \over A}  |\tau + \epsilon |^A \sin{\pi A}\cr
&\hskip10pt  + {4\alpha' G \over A} |\tau - \epsilon |^A \cos{\pi A}
+  {2\alpha' G B \over A}  |\tau - \epsilon |^A \sin{\pi A}\cr  
&=  -4\alpha'\pi (g)^2 {q_2\pi\alpha' B_{12}\over
{1 +  (g)^2 (q_2 2\pi\alpha' B_{12})^2}}\cr
&= 2 \tilde\Theta ^{12}\,.\cr}}

In the limit $q_1\rightarrow -q_2$, then 
$ \tilde\Theta ^{12} \rightarrow - \Theta ^{12}$.
Indeed
in the neutral string case, 
where both ends of the string are on the same $D$-brane,
the noncommutativity parameter at one end of the string is
equal to minus that of the other end. 
For $U(N)$ Chan-Paton factors, the background magnetic fields 
can take on N possible values, giving rise to $N$ noncommutativity
parameters. 

We remark that had we computed the charged string propagator 
between the states $\langle x_-| $ and $|x_+\rangle$, 
rather that $\langle\beta|$, $|\alpha\rangle$, we would have
found the same noncommutativity parameters as derived in 
\scho\ and \schpi\ , but the $A\rightarrow 0$ limit of this propagator
would be different from the neutral string expression. 
\vfill\eject

\newsec{Short Distance Behavior and Star Products}
\def\half{{1\over 2}}


For $\tau > \tau'$ and at $\sigma = 0$,
the leading short distance singularity in the product of tachyon vertex
operators is 
\eqn\sdb{\eqalign{&  e^{ip\cdot X(\tau)}   \,  e^{iq\cdot X(\tau')}  \,=\,
: e^{ip_- X^+(\tau) + ip_+ X^-(\tau)} : \, 
: e^{iq_- X^+(\tau') + iq_+ X^-(\tau')} :\cr
&\sim e^{ip_- X^+_< (\tau) + ip_+ X^-_< (\tau)} 
\,e^{ip_- X^+_> (\tau) + ip_+ X^-_> (\tau)} 
\, e^{ip_- x^+ + ip_+ x^-}\cr
&\hskip 20pt \cdot \, e^{iq_- x^+ + iq_+ x^-}\, 
e^{iq_- X^+_< (\tau') + iq_+ X^-_< (\tau')}\, 
e^{iq_- X^+_> (\tau') + iq_+ X^-_> (\tau')}\cr
&\sim  e^{-4\alpha' G p_- q_+ \sum_{n\ge 0} {1\over {n+A}}
({\tau'\over{\tau}})^{(n+A)}} \,
e^{-4\alpha' G p_+ q_- \sum_{n>0} {1\over {n-A}}
({\tau'\over{\tau}})^{(n-A)}}\cr 
&\hskip 20pt \cdot \, e^{-\half (p_- q_+ - p_+ q_-)\, 
({-4g\alpha'\pi\over{B+{q_2\over q_1}B}})}\,
: e^{i (p_- + q_-) X^+(\tau') + i (p_+ + q_+) X^-(\tau')} :\cr
&\sim  e^{(p_- q_+  + p_+ q_- ) (-2\alpha' G)
(-2 \ln (\tau -\tau')}\cr
&\hskip 20pt \cdot \,
e^{ -\Theta^{12} (p_- q_+  - p_+ q_- )} \,
: e^{i (p_- + q_-) X^+(\tau') + i (p_+ + q_+) X^-(\tau')} :\cr
&\sim (\tau -\tau')^{4\alpha' G\,(p_- q_+  + p_+ q_- )}\,\,
e^{ -\Theta^{12} (p_- q_+  - p_+ q_- )} \,
: e^{i (p + q) \cdot X(\tau')}:\cr}}
where we have defined
\eqn\scal{\eqalign{X^+_{>\atop <} (\tau)
&\equiv i\sqrt{2\alpha'} \sum_{n{\ge\atop <}0}
{a_{n+A}\over {n+A}} \tau^{-n-A};\qquad
X^-_{>\atop <} (\tau)
\equiv i\sqrt{2\alpha'}
\sum_{n{>\atop\le}0}{\tilde a_{n-A}\over {n-A}} \tau^{-n+A}\,.\cr}}

In order to derive the next to the last line in \sdb\
we need to use the same identities used in computing \scho. 
Note that \sdb\ holds for fixed $A$.
In the limit scaling limit \sw\ , ($\alpha'\rightarrow 0$,
keeping $G$ and $\Theta^{12}$ fixed), the OPE 
reduces to the star product.
\eqn\scaltach{\eqalign{ e^{ip\cdot X(\tau)}   \,  e^{iq\cdot X(\tau')}
&\sim e^{ -\Theta^{12} (p_- q_+  - p_+ q_- )} \,
: e^{i (p + q) \cdot X(\tau')} :\cr
&\equiv  e^{ip\cdot X(\tau')} \ast  e^{iq\cdot X(\tau')}\,.\cr}}
For $\sigma = \pi$ the same equations \sdb ,\scaltach ,
will hold with $\Theta^{12}$
replaced by $\tilde\Theta^{12}$.

To accomodate the existence of many noncommutativity parameters,
the non-abelian $U(N)$ gauge theory (whose expansion around a 
$U(1)^N$ background we are considering here), must require
a generalization of the usual noncommutative star product. 
It would be interesting to figure out the
Lagrangian for our noncommutative gauge theory as well as the
underlying $U(N)$ gauge transformations.


\newsec{The Spectrum in the Scaling Limit}

As in \sw\ , we consider the scaling limit
$g^{-1}\rightarrow \epsilon $ and
$\alpha'\rightarrow \sqrt{\epsilon}$, for $\epsilon\rightarrow 0$.
The other quantities $B_{12}, q_1, q_2$ are held fixed. 
Actually this limit means letting the dimensionless quantity
$\alpha' B_{12} \rightarrow \sqrt{\epsilon}$, while keeping $B_{12}$ fixed. 
Then
we have that the noncommutativity parameters are
finite in the scaling limit and are given by
$\Theta^{12}\rightarrow (q_1 B_{12} )^{-1}$,
$\tilde \Theta^{12}\rightarrow (q_2 B_{12} )^{-1}$.

Starting from \somedef\ , we have 
$\tan {\pi A} = {{B + {q_2\over q_1} B}\over  
{1 - {q_2\over q_1} B^2}}$. So in the scaling limit,
\eqn\tlim{\tan {\pi A} \rightarrow -{{(q_1+q_2) {\sqrt{\epsilon}}}\over
2\pi q_1 q_2 }\,.} 
and
\eqn\masslim{ A\rightarrow -{{(q_1+q_2) {\sqrt{\epsilon}}}\over
2 \pi^2 q_1 q_2 }\,.}
Therefore the mass formulae for the two polarization states of the 
charged vector
we have listed in \tower\ are also finite in the scaling limit 
and behave as 
\eqn\mtach{\eqalign{{\rm mass}^2 &
=-{1\over 2\alpha'} A (1+A) \rightarrow {1\over 2}{{(q_1+q_2) B_{12}}
\over 2 \pi^2 q_1 q_2 } = 
{(q_1+q_2) G
\over q_2 \Theta^{12}} \,,\cr
{\rm mass}^2 & 
={1\over 2\alpha'} A (3-A))
\rightarrow -{3\over 2}{(q_1+q_2) B_{12}\over
2\pi^2 q_1 q_2 } =
-3 {(q_1+q_2) G
\over q_2 \Theta^{12}}\,.\cr}}

Also in the spectrum
are the charged vectors at different Landau levels. 
For each charged boson, the two polarizations have different masses.
They differ from those of \mtach\ by integer multiples
of ${2(q_1+q_2)G\over q_2 \Theta^{12}}$. 
(Notice that each Landau level has infinite degeneracy labeled by 
$|x_+\rangle$). 
All other states in the charged string spectrum become infinitely
heavy and decouple in the scaling limit.
So the complete spectrum of our $U(1)^N$ noncommutative field theory,
which is derived from $N$ 
neutral and $N^2 - N$ charged string sectors, is described
by $N$ massless neutral gluons and the charged vectors above.

We have shown that our theory, which is a $U(N)$ gauge theory expanded 
around a $U(1)^N$ background, has a scaling limit.
It would be of interest to work out the scaling limit of the full
non-abelian gauge theory and its noncommutativity parameters.

\newsec{Acknowledgements}

CRN thanks the Institute for Advanced Study and Princeton University for
hospitality during the summer 2000, 
and was partially supported by the  U.S. Department of Energy,
Grant No. DE-FG 03-84ER40168.
LD thanks the CIT-USC Center for Theoretical Physics, the
IAS, and the Aspen Center for Physics for their hospitality.
Research of LD was supported in part by U.S. Department of Energy,
Grant No. DE-FG 05-85ER40219/Task A.

\listrefs

\bye